\documentclass[prc,twocolumn,showpacs,floatfix,superscriptaddress,nofootinbib]{revtex4}
\usepackage{graphicx}
\usepackage{amssymb}
\usepackage{dcolumn}
\usepackage{amsmath}
\usepackage{bm}
\usepackage{textcomp}
\usepackage{epstopdf}
\usepackage{color}
\usepackage{ulem}
\usepackage[toc,page,title,titletoc,header]{appendix}
\usepackage[colorlinks,
            linkcolor=blue,
            anchorcolor=blue,
            citecolor=blue
            ]{hyperref}
\usepackage{txfonts}
\usepackage{caption}

\begin{document}

\title{Universal scaling of the $\sigma$ field and net-protons from Langevin dynamics of model A}

\begin{abstract}
In this paper, we investigate the Kibble-Zurek scaling of the $\sigma$ field and net-protons within the framework of Langevin dynamics of model A. After determining the characteristic scales $\tau_{\mbox{\tiny KZ}},l_{\mbox{\tiny KZ}}$ and $\theta_{\mbox{\tiny KZ}}$ and properly rescaling the traditional cumulants, {we construct universal functions for the $\sigma$ field and approximate universal functions  for net-protons in the critical regime, which are insensitive to the relaxation time and the chosen evolving trajectory}. Besides, the oscillating behavior for the higher order cumulants of net-protons near the critical point is also drastically suppressed, which converge into approximate universal curves with these constructed Kibble-Zurek functions.
\end{abstract}

\author{Shanjin Wu}
\email{shanjinwu2014@pku.edu.cn}
\affiliation{Department of Physics and State Key Laboratory of Nuclear Physics and
Technology, Peking University, Beijing 100871, China}

\author{Zeming Wu}
\email{zeming$_$wu@pku.edu.cn}
\affiliation{Department of Physics and State Key Laboratory of Nuclear Physics and
Technology, Peking University, Beijing 100871, China}

\author{Huichao Song}
\email{huichaosong@pku.edu.cn}
\affiliation{Department of Physics and State Key Laboratory of Nuclear Physics and
Technology, Peking University, Beijing 100871, China}
\affiliation{Collaborative Innovation Center of Quantum Matter, Beijing 100871, China}
\affiliation{Center for High Energy Physics, Peking University, Beijing 100871, China}
\maketitle


\section{Introduction}\label{Intro}

The search for the critical point on the phase diagram of quantum chromodynamics (QCD) has attracted considerable attention in the heavy ion community for decades~\cite{Stephanov:1998dy,Aggarwal:2010cw,Stephanov:2004wx,Stephanov:2007fk,Asakawa:2015ybt,Luo:2017faz}. The critical point is the endpoint of the first order phase transition boundary that separates the quark-gluon plasma phase and the hadronic phase~\cite{Stephanov:1998dy,Stephanov:2004wx,Berges:2000ew,Qin:2010nq,Roberts:1994dr,Jiang:2013yoa,Fu:2007xc,Fukushima:2003fw,Klevansky:1992qe}. The characteristic features of the critical point are the divergence of various fluctuations, long range correlations and singularities of some thermodynamic quantities~\cite{Stephanov:2004wx}.  For example, the variance $\sigma$, skewness $S$ and kurtosis $\kappa$ of the $\sigma$ field are proportional to various orders of the correlation length $\xi$, which diverge with $\xi^2$~\cite{Stephanov:1998dy,Stephanov:1999zu}, $\xi^{4.5}$ and $\xi^{7}$, respectively~\cite{Stephanov:2008qz}. It was also found that the kurtosis $\kappa$ of the $\sigma$ field presents a non-monotonic behavior with the increase of the net-baryon chemical potential $\mu_B$~\cite{Stephanov:2011pb}. After
coupling the $\sigma$ field with various hadrons, such critical fluctuations also influence the multiplicity distributions of conserved charges~\cite{Stephanov:2008qz}, which can be systematically measured in experiment.

The Beam Energy Scan (BES) program at BNL Relativistic Heavy-Ion Collider (RHIC) aims  to search the QCD critical point through evaluating the fluctuations of conserved charges~\cite{Aggarwal:2010wy,Adamczyk:2013dal,Luo:2015ewa,Adamczyk:2014fia,Thader:2016gpa}. Recently, higher order cumulants of net-protons, with the transverse momentum coverage extended to $0.4<p_T<2$ GeV, have been systematically measured in Au+Au collisions from 7.7 to 200 $A$ GeV~\cite{Luo:2015ewa}. The kurtosis of net-protons in the most central collisions presents a non-monotonic behavior and largely  deviates from the poisson baseline below 39 GeV, which indicates the potential of discovering the critical point.

In the theoretical side, the equilibrium and nonequilibrium critical fluctuations near the critical point have been investigated by different groups~\cite{Stephanov:2008qz,Stephanov:2011pb,Stephanov:2009ra,Son:2004iv,Ling:2015yau,Stephanov:2017ghc,Mukherjee:2015swa, Mukherjee:2016kyu,Brewer:2018abr,Akamatsu:2018vjr,Rajagopal:2000,Paech:2003fe,Nahrgang:2011mg,Nahrgang:2018afz, Nonaka:2004pg,Asakawa:2009aj,Sakaida:2017rtj,Jiang:2015hri,Jiang:2017mji}.
Through coupling the order parameter field to the emitted protons and anti-protons on the freeze-out surface, the equilibrium critical fluctuations qualitatively explained the acceptance dependence of the measured cumulants and the non-monotonic behavior of the kurtosis for net-protons~\cite{Jiang:2015hri,Ling:2015yau}. However, the same framework failed to describe the cumulants $C_2$ and $C_3$ of net-protons due to the intrinsic positive contributions of the equilibrium critical fluctuations~\cite{Jiang:2015hri}. Recently, it was realized that the critical slowing down effects largely influence the nonequilibrium critical fluctuations, which even reverse the signs of skewness and kurtosis compared to the equilibrium values~\cite{Mukherjee:2015swa,Jiang:2017mji}. Besides, it was also found that the nonequilibrium evolution near the critical point also influences the rapidity window dependence of the variance~\cite{Sakaida:2017rtj}. For a qualitative and quantitative evaluation of the BES data and for the search of the  critical point,
it is important to develop dynamical models for the evolving bulk matter together with nonequilibrium evolution of the critical mode (For recent progresses, please also refer to the work of hydro+~\cite{Stephanov:2017ghc}).

For a dynamical model near the critical point, the calculated nonequilibrium fluctuations are sensitive to various free inputs and parameters, such as the trajectory and relaxation time of the evolving system, the mapping between the three-dimensional Ising model and the hot QCD system, etc.  Meanwhile, the critical slowing down effects drive the system out of equilibrium, which leads to correlated regions with characteristic length scales after the system becomes ``frozen''. It was realized that,  within the framework of Kibble-Zurek mechanism (KZM), one could construct some universal variables near the critical point, that are independent on some of these non-universal factors~\cite{Chandran:2012,Kolodrubetz:2012,Francuz:2015zva,Nikoghosyan:2013fqa,Mukherjee:2016kyu}.
In cosmology, the KZM was first introduced by Kibble~\cite{Kibble:1976}  to study the defect formation of the expanding Universe after the Big Bang, which is then extended by Zurek~\cite{Zurek:1985} to study the condensed matter systems near the critical point. {Recently, the KZM was applied to relativistic heavy ion collision within the framework of Fokker-Planck equation, which constructed universal functions for the evolving $\sigma$ field  in the critical
regime~\cite{Mukherjee:2016kyu}.

In this work, we will investigate the universal scaling of both $\sigma$ field and net-protons from the Langevin dynamics of model A.  Compared with the Fokker-Planck equation approach which only considers the zero mode of the $\sigma$ field, our Langevin dynamics simulations  evolve the whole $\sigma$ field in the position space event by event, which can be coupled with final hadrons to further investigate the multiplicity fluctuations and possible universal scaling of net-protons in the critical regime. Note that this paper does not aim to construct realistic universal experimental observables with this simplified  Langevin dynamics,  but focuses on investigating the Kibble-Zurek scaling of the $\sigma$ field and net-protons with two ideal cases: 1) the systems evolve along a chosen trajectory with different relaxation times; 2) the systems evolve along different chosen trajectories. We will demonstrate that one could construct universal functions for the $\sigma$ field in the critical regime, which are insensitive to the relaxation time or chosen evolution trajectory. With a linear expansion of the distribution functions of protons and anti-protons, such universal behavior of the $\sigma$ field could be translated into a similar universal behavior of net-protons through the $\sigma N N$ coupling. On the other hand, the numerical simulations with the full distribution functions of protons and anti-protons show that one could still construct an approximate universal functions for net-protons, which drastically reduce the sensitivity to the relaxation time and evolving trajectory.

{The paper is organized as follows.  Section~\ref{Model} briefly introduces Langevin dynamics of model A and the basic idea to construct the universal functions according to the  Kibble-Zurek mechanism. Section~\ref{Results} presents and discusses the constructed universal functions of the $\sigma$ field and approximate universal functions of net-protons in the critical regime. Section~\ref{summary} summarizes and concludes the paper.}

\section{Model and set ups}\label{Model}
\subsection{Langevin dynamics of model A}\label{ModelLangevin}

It is generally believed that the hot QCD system belongs to model H according to the classification of Ref.~\cite{Rev1977}, which focuses on the dynamics of order parameter field, baryon and energy density~\cite{Son:2004iv}. Recently, an alternative approach, hydro+~\cite{Stephanov:2017ghc}, has been developed,  which extended traditional hydrodynamics to the critical regime with the additional evolution of the slow mode. However the numerical implementation of model H and hydro+ are both complicated, which are still under development.

In this paper, we focus on investigating the universal behavior of the $\sigma$ field and net-protons near the critical point, with a simplified  Langevin dynamics, called model A, that only evolves the non-conserved order parameter field of one single component. The corresponding equation is written as:

\begin{align}\label{Langevin3+1}
  \frac{\partial \sigma(\bm{x},\tau)}{\partial \tau} = - \frac{1}{m^2_\sigma \tau_{\mbox{\tiny eff}}} \frac{\delta U[\sigma(\bm{x})]}{\delta \sigma(\bm{x})} + \zeta(\bm{x},\tau),
\end{align}
where the noise $\zeta$ satisfies the fluctuation-dissipation theorem:
\begin{align}\label{noise3+1}
\begin{aligned}
  &\langle \zeta(\bm{x},\tau) \rangle = 0,\\
  &\langle \zeta(\bm{x},\tau) \zeta(\bm{x}',\tau')\rangle = \frac{2T}{m^2_\sigma \tau_{\mbox{\tiny eff}}} \delta^3(\bm{x}-\bm{x}') \delta(\tau-\tau'),
\end{aligned}
\end{align}
Here, $T$ is the temperature, $m_\sigma$ is the mass of the $\sigma$ field, $\tau_{\mbox{\tiny eff}}$ is the relaxation time and $U[\sigma(\bm{x})]$ is the effective potential. According to the analyses of dynamical critical behavior~\cite{Rev1977}, { the effective relaxation time $\tau_{\mbox{\tiny eff}}$ depends on equilibrium correlation length $\xi_{\mbox{\tiny eq}}$ as $\tau_{\mbox{\tiny eff}}= \tau_{\mbox{\tiny rel}}(\xi_{\mbox{\tiny eq}}/\xi_{\mbox{\tiny min}})^z$},  where $\tau_{\mbox{\tiny rel}}$ is a free parameter in this work. For the dynamical critical exponent, we use the one from model H with $z = 3$.

In the vicinity of the critical point, the effective potential $U[\sigma(\bm{x})]$ can be expanded in the powers of the order parameter field $\sigma(\bm{x})$:
\begin{align}\label{effaction}
\begin{aligned}
  U[\sigma(\bm{x})]= \int &d^3x\left\{\frac{1}{2}[\nabla\sigma(\bm{x})]^2+\frac{1}{2}m^2_\sigma[\sigma(\bm{x})-\sigma_0]^2 \right.\\
  &\left.+\frac{\lambda_3}{3}[\sigma(\bm{x})-\sigma_0]^3+\frac{\lambda_4}{4}[\sigma(\bm{x})-\sigma_0]^4\right\},
\end{aligned}
\end{align}
where  $\lambda_3$ and $\lambda_4$ are the coupling coefficients of the cubic and quadratic terms, $\sigma_0$ is the equilibrium mean value of $\sigma(\bm{x})$, $m_\sigma$ is the mass of the $\sigma$ field which is related to the equilibrium correlation length with $m_\sigma=1/\xi_{eq}$. Following Ref.~\cite{Mukherjee:2015swa}, we construct the effective potential  $U[\sigma(\bm{x})]$ through a mapping between the hot QCD system and the three-dimensional Ising model~\cite{Justin:2001,Schofield:1969}. In more details,  one first calculates the cumulants from the distribution function $P[\sigma]\sim \exp[-U(\sigma)/T]$ and from the parametrization of magnetization $M^{\mbox{\tiny eq}}$ of the 3d Ising model.  A comparison of  the cumulants obtained from these two procedures gives the forms of $\sigma_0(R,\theta),\xi(R,\theta),\lambda_3(R,\theta)$ and $\lambda_4(R,\theta)$. Here, $R$ and $\theta$ are the distance and angle with respect to the location of the critical point, which are related to the Ising model variables $r$ and $h$ via: $ r(R,\theta)= R(1-\theta^2),\, h(R,\theta) = R^{5/3}(3\theta-2\theta^3)$. {The Ising model variables $(r,\ h)$ are related to the hot QCD parameters $(T,\ \mu)$  through a linear mapping :  $(T-T_c)/\Delta T = h/\Delta h,\, \ (\mu-\mu_c)/\Delta \mu = - r/\Delta r$. Note that such mapping is non-universal, which depends on the position of the critical point and the shape of the critical regime for the constructed QCD phase diagram. The details can be found in the Appendix of this paper.

To numerically solve Eq.~(\ref{Langevin3+1}), one needs to input the local temperature $T(\bm{x})$ and
local chemical potential $\mu(\bm{x})$ of the external heat bath. For simplicity, we assume that the heat bath evolves along certain trajectory with uniform temperature and chemical potential in the position space. Such trajectories can be expressed with the $r$ and $h$ variables~\cite{Mukherjee:2016kyu}:
\begin{align}
  r=r_c-a_h h^2,
\end{align}
where $r_c$ and $a_h$ are two free parameters to tune the shape of the trajectories. In the following calculations, we select two types of trajectories, called type A and type B. For type A trajectory, we set $a_h=0$ and $r_c=0.02\ \Delta r$. This corresponds to the system evolving with fixed chemical potential, where the changing rate of the effective potential is mainly captured by the variation of the correlation length $\xi_{\mbox{\tiny eq}}$.  For type B trajectories, we set $r_c$ as a constant, and tune $a_h$ to ensure approximately equal correlation length near the phase transition line. In this way, $\theta$ becomes the dominated factor for the changing rate of the effective potential, {which simplifies the corresponding  analysis of the  universal behavior for type B trajectories}. In Fig.~\ref{Traj}, we plot the trajectories of type A and type B, which are denoted by the magenta line and different colored curves, respectively. The black dashed curve is  the boundary of the critical regime defined by the equilibrium correlation length $\xi_{\mbox{\tiny min}}=1$ fm.  {Note that, a more realistic trajectory of the heat bath is the line with constant baryon density over entropy $n/s$, which is complicate for the following computation of quench time in Sec.~\ref{Univ}. For simplicity, we assume the system evolves along these chosen trajectories (type A and type B) in this work.}

As the heat bath evolving along one of these trajectories, we assume that the temperature $T$ drops down in a Hubble-like way~\cite{Mukherjee:2016kyu}:
\begin{equation}\label{Hubble}
\frac{T\left( \tau\right) }{T_{I}}=\left( \frac{\tau}{\tau_{I}}\right) ^{-0.45},
\end{equation}%
where $T_{I}$  and $\tau_{I}$  are the initial temperature and initial time.

In numerical simulations, we first construct the initial profiles of the $\sigma$ field through the probability function: $ P\left[ \sigma(\bm{x}) \right] \sim \exp \left\{ -U \left[ \sigma(\bm{x})\right] /T\right\} $ and  then evolve the $\sigma$ field event by event according to Eq.~(\ref{Langevin3+1}). But the discretization of the noise term  leads to a grid size dependence for the calculated cumulants~\cite{Cassol-Seewald:2012}. To avoid this complexity, we only focus on the long wavelength behavior of the evolving system and coarse-grain the noise term over the spatial extension as proposed in Ref.~\cite{Herold:2016uvv}.  In other word, we numerically evolve Eq.~(\ref{Langevin3+1}) in 3+1 dimensions and the noise term is been coarse-grained which is uniform in coordinate space but random in temporal direction.  In the limit of zero mode, one can prove~\cite{Wu:2018note} that such Langevin equation is equivalent to the Fokker-Planck equation implemented in Ref.~\cite{Mukherjee:2015swa}.
For each time step, we calculate the corresponding cumulants which are defined as the following:
\begin{equation}
\begin{split}\label{cumulants}
C_{1}&=\langle \sigma \rangle,   \\
C_{2}&=\langle \sigma^2 \rangle - \langle \sigma \rangle^2,  \\
C_{3}&=\langle \sigma ^3 \rangle -3\langle \sigma^2 \rangle \langle \sigma \rangle + 2\langle \sigma \rangle^3,   \\
C_{4}&=\langle \sigma^4 \rangle -4\langle \sigma^3 \rangle \langle \sigma \rangle -3\langle \sigma^2 \rangle^2 +12 \langle \sigma^2\rangle\langle \sigma \rangle^2 -6\langle \sigma \rangle^4,
\end{split}
\end{equation}
where $\sigma$ denotes the spatial average of the $\sigma$ field $\sigma(\bm{x})$ and $\langle \cdots\rangle$ is the event average.

\begin{figure}[tbp]
\center
\includegraphics[width=3.0 in]{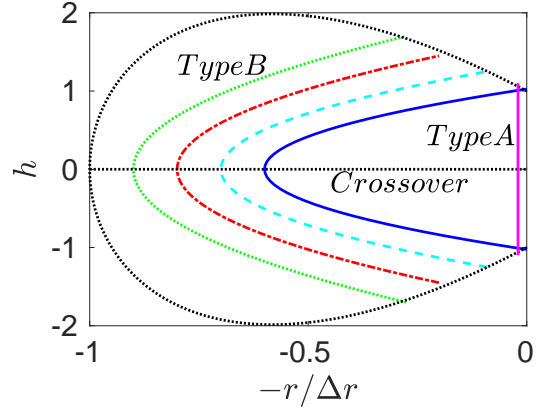}
\caption{(Color online)  Trajectories of type A (magenta line) and Type B (colored curves) in the critical regime. The boundary of the critical regime (black dashed curve) is defined by $\xi_{eq}=1$ fm.}
\label{Traj}
\end{figure}
\begin{figure*}[tbp]
\center
\includegraphics[width=3.3 in]{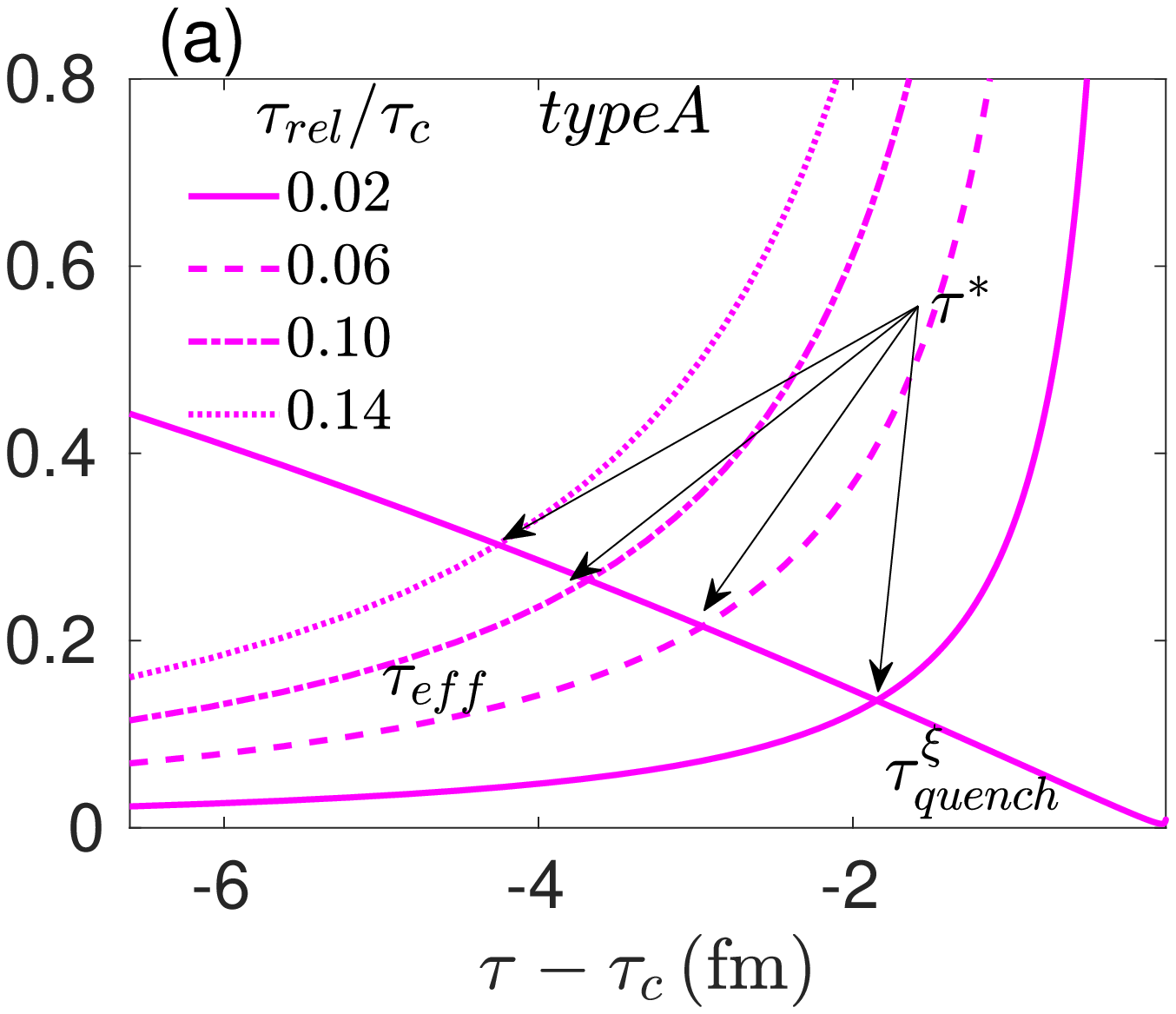}
\includegraphics[width=3.3 in]{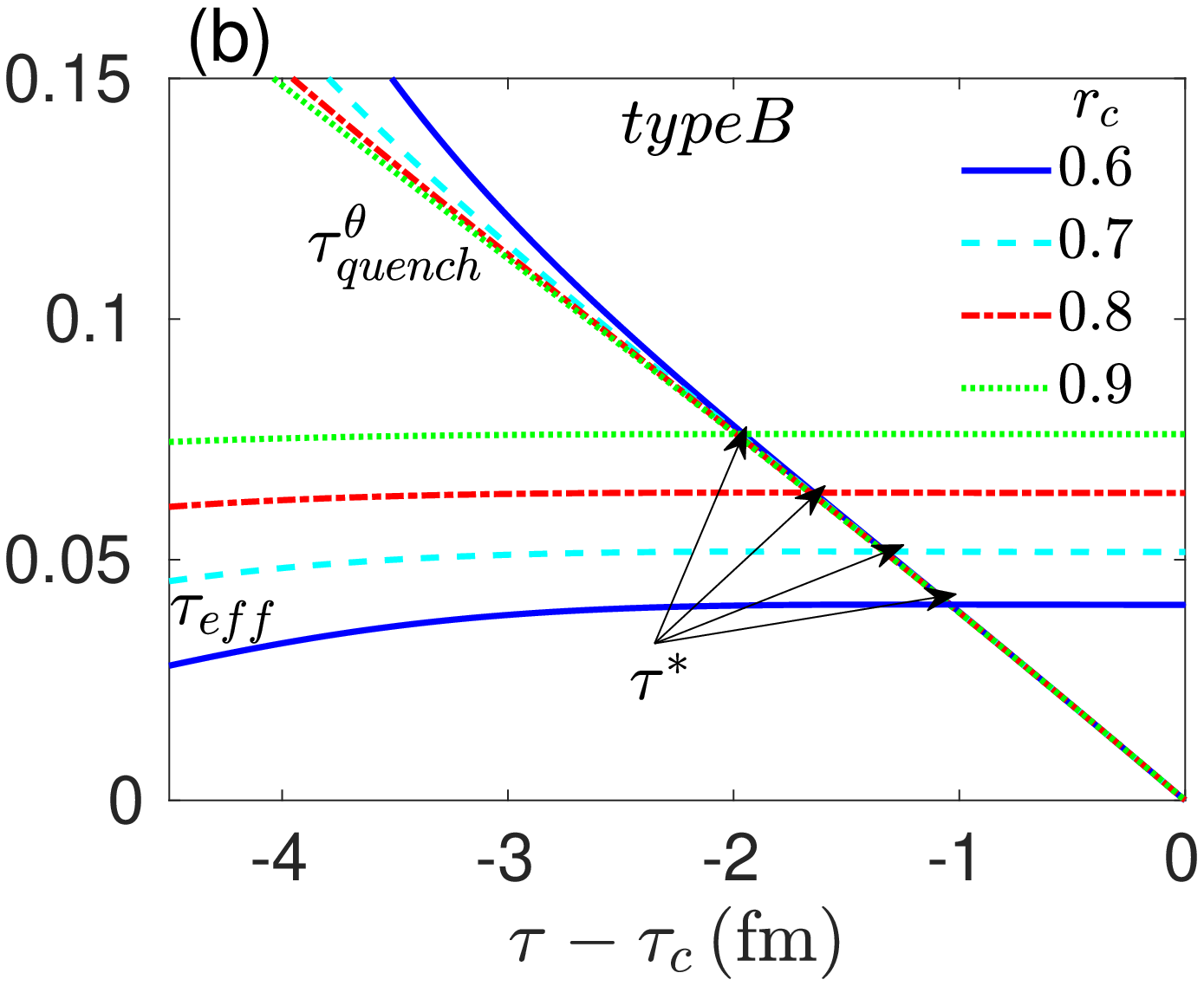}
\caption{(Color online) Time evolution of the relaxation time $\tau_{\mbox{\tiny eff}}$  and quench time $\tau_{\mbox{\tiny quench}}$ along trajectory of type A with different $\tau_{rel}/\tau_{c}$ (a) and along trajectories of type B with different $r_c$ (b). The locations of the proper time $\tau^*$ are obtained from $\tau_{\mbox{\tiny eff}}(\tau^*) = \tau_{\mbox{\tiny quench}} (\tau^*)$.}
\label{xiKZ}
\end{figure*}

\subsection{{The Kibble-Zurek scaling}}\label{Univ}
The above cumulants of the $\sigma$ field Eq.~(\ref{cumulants}) are influenced by inputs and free parameters in the model calculations, such as the relaxation time, the trajectory of the heat bath, and the mapping between the 3d Ising model and the hot QCD system, etc. Within the framework of the Kibble-Zurek mechanism, Ref.~\cite{Mukherjee:2016kyu} has constructed some universal functions for the Fokker-Planck equation approach, which are independent on some non-universal factors. In this paper, we will explore such universal behavior within the framework of Langevin dynamics.

{For a system evolving near the critical point, there are two competitive time scales, the relaxation time $\tau_{\mbox{\tiny eff}}$ that describes the relaxation rate of the order parameter field and the quench time $\tau_{\mbox{\tiny quench}}$ that describes the changing rate of the effective potential}. As explained in Sec.~\ref{ModelLangevin}, the relaxation time takes the form  $\tau_{\mbox{\tiny eff}}= \tau_{\mbox{\tiny rel}}(\xi_{\mbox{\tiny eq}}/\xi_{\mbox{\tiny min}})^z$ with $z=3$. The quench time $\tau_{\mbox{\tiny quench}}=\mbox{min}\left(\tau^\xi_{\mbox{\tiny quench}},\tau^\theta_{\mbox{\tiny quench}}\right)$ can be calculated
as~\cite{Mukherjee:2016kyu}:
\begin{align}
  \tau^\xi_{\mbox{\tiny quench}} = \left|\frac{\xi_{eq}(\tau)}{\partial_\tau\xi_{eq}(\tau)}\right|,\quad \tau^\theta_{\mbox{\tiny quench}} =\left|\frac{\theta(\tau)}{\partial_\tau \theta(\tau)}\right|.
\end{align}
In general, the quench time decreases as the system cools down, and the relaxation time rapidly increases as the system approaches the critical point due to the critical slowing down effects.  This leads to a point $\tau^*$, where the relaxation time equals to the quench time, after which the order parameter field becomes hard to adjust itself to the changing effective potential. In other words, the system becomes approximately frozen after $\tau^*$.  Correspondingly, one defines the characteristic time scale $\tau_{\mbox{\tiny KZ}}$, length scale $l_{\mbox{\tiny KZ}}$ and magnetization angle $\theta_{\mbox{\tiny KZ}}$ to characterize the typical scales of the correlated patches for the evolving systems near the critical point:
\begin{align}\label{KZ}
\begin{aligned}
  \tau_{\mbox{\tiny KZ}} = \tau_{\mbox{\tiny eff}}(\tau^*) = \tau_{\mbox{\tiny quench}} (\tau^*),\\
  l_{\mbox{\tiny KZ}} = \xi_{\mbox{\tiny eq}}(\tau^*),\quad \theta_{\mbox{\tiny KZ}} = \theta(\tau^*).
\end{aligned}
\end{align}

In Fig.~\ref{xiKZ}, we plot the time evolution of the relaxation time $\tau_{\mbox{\tiny eff}}$  and quench time $\tau_{\mbox{\tiny quench}}$ along trajectories of type A (with $\tau_{rel}/\tau_c =0.02,\,0.06,\,0.10,\,0.14 $) and type B (with $r_c=0.6,\,0.7,\,0.8,\,0.9$). As shown in Fig.~\ref{xiKZ},  the increasing relaxation time $\tau_{\mbox{\tiny eff}}$ and decreasing quench time $\tau_{\mbox{\tiny quench}}$ lead to a proper time $\tau^*$, with which one could further obtain the characteristic scales $\tau_{\mbox{\tiny KZ}},l_{\mbox{\tiny KZ}}$ and $\theta_{\mbox{\tiny KZ}}$ from  Eq.~(\ref{KZ}).} {Following~\cite{Mukherjee:2016kyu}, we construct the universal functions $\bar{f}_{n}((\tau-\tau_{\mbox{\tiny KZ}})/\tau_{\mbox{\tiny KZ}},\theta_{\mbox{\tiny KZ}}) \ (n=1,\dots,4)$ through rescaling
the cumulants $C_n \ (n=1,\dots,4)$ and the proper time $\tau-\tau_c$ with these characteristic scales, which is written as the following:}

\begin{align}\label{scalansatz}
  C_n(\tau-\tau_c) \sim l^{-\frac{1}{2}+\frac{5}{2}(n-1)}_{\mbox{\tiny KZ}} \bar{f}_n[(\tau-\tau_c)/\tau_{\mbox{\tiny KZ}};\theta_{\mbox{\tiny KZ}}],  \  \ \  \ n=1,\dots,4
\end{align}
where  $\tau_c$ is the time when the QCD system evolves to the crossover line.
The exponent of $l_{\mbox{\tiny KZ}}$ comes from the fact that the $n$-order equilibrium critical cumulants are proportional to $[-1+5(n-1)]/2$ powers of the correlation length $\xi_{\mbox{\tiny eq}}$~\cite{Mukherjee:2016kyu}.

\section{Results and discussions}\label{Results}
In the following calculations, we first simulate the evolution of the $\sigma$ field using  Eq.~(\ref{Langevin3+1}), and then investigate the possible universal behavior of the $\sigma$ field and net-protons. Below are detailed calculations and results.

\subsection{Kibble-Zurek scaling of the $\sigma$ field}

Fig.~\ref{A} (a) shows the time evolution of the cumulants for the sigma field,
which evolves along a fixed trajectory of type A with different relaxation times, $\tau_{\mbox{\tiny rel}}/\tau_c=0.02,0.06,0.1,0.14$. Due to the critical slowing down effects,  these dynamical
cumulants largely deviate from the equilibrium values, which are also sensitive to the relaxation time inputs.  Fig.~\ref{A} (b) focuses on demonstrating the related
universal behavior of the sigma field. As explained in Sec.~\ref{Model}, the changing rate of the effective potential along type A trajectory  is mainly controlled by the variance of the correlation length $\xi_{eq}$, and the corresponding quench time is: $\tau^\xi_{\mbox{\tiny quench}} = \left|\xi_{eq}(\tau)/\partial_\tau\xi_{eq}(\tau)\right|$.
{As shown in Fig.~\ref{xiKZ} (a), the proper time $\tau^*$ can be obtained from comparing the relaxation time $\tau_{\mbox{\tiny eff}}$ and the quench time  $\tau^\xi_{\mbox{\tiny quench}}$, with which the Kibble-Zurek
scales $\tau_{\mbox{\tiny KZ}}$ and $l_{\mbox{\tiny KZ}}$ can be calculated from Eq.~(\ref{KZ})}.   With  $l_{\mbox{\tiny KZ}}$ and $\tau_{\mbox{\tiny KZ}}$, we rescale $C_n$ and $\tau-\tau_c$  and construct the universal functions $\bar{f}_n$  according to Eq.~(\ref{scalansatz}).  Fig.~\ref{A} (b) plots the universal functions $\bar{f}_n$ for the evolving systems with different relaxation times, which converge into one universal curve near the critical point. In contrast, the original cumulants, $C_n(n=1,\dots,4)$, before the rescaling procedure are separated from each other and are sensitive to the relaxation times.
\begin{figure*}[tbp]
\center
\hspace{5mm}
\includegraphics[width=7.3 in,height=2.9in]{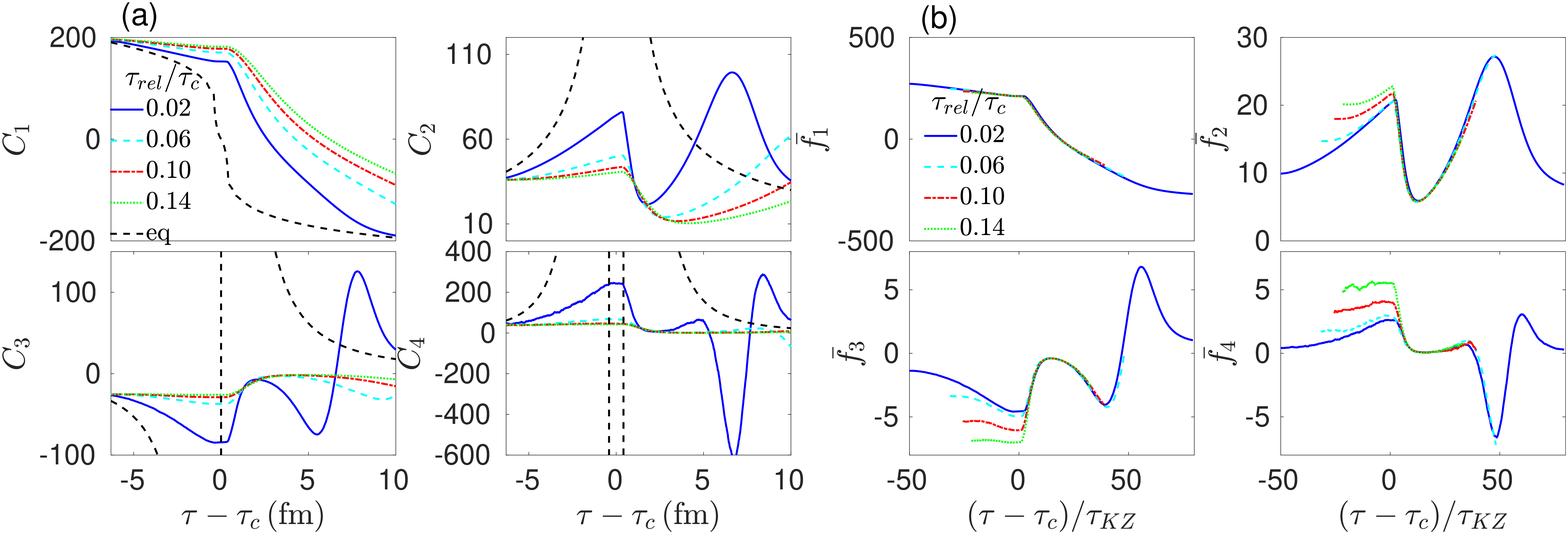}
\caption{(Color online){ (a): The cumulants $C_n(n=1,\dots,4)$ of the sigma field as a function of  $\tau-\tau_c$, evolving along type A trajectory with $\tau_{\mbox{\tiny rel}}/\tau_c=0.02,\ 0.06, \ 0.1,\ 0.14$. The dashed curves represent the equilibrium cumulants along the trajectory. (b): The corresponding universal functions $\bar{f}_n((\tau-\tau_{\mbox{\tiny c}})/\tau_{\mbox{\tiny KZ}},\theta_{\mbox{\tiny KZ}})~(n=1,\dots,4)$ as a function of  $(\tau-\tau_c)/\tau_{KZ}$.}}
\label{A}
\end{figure*}

\begin{figure*}[tbp]
\center
\includegraphics[width=7.3 in,height=2.9in]{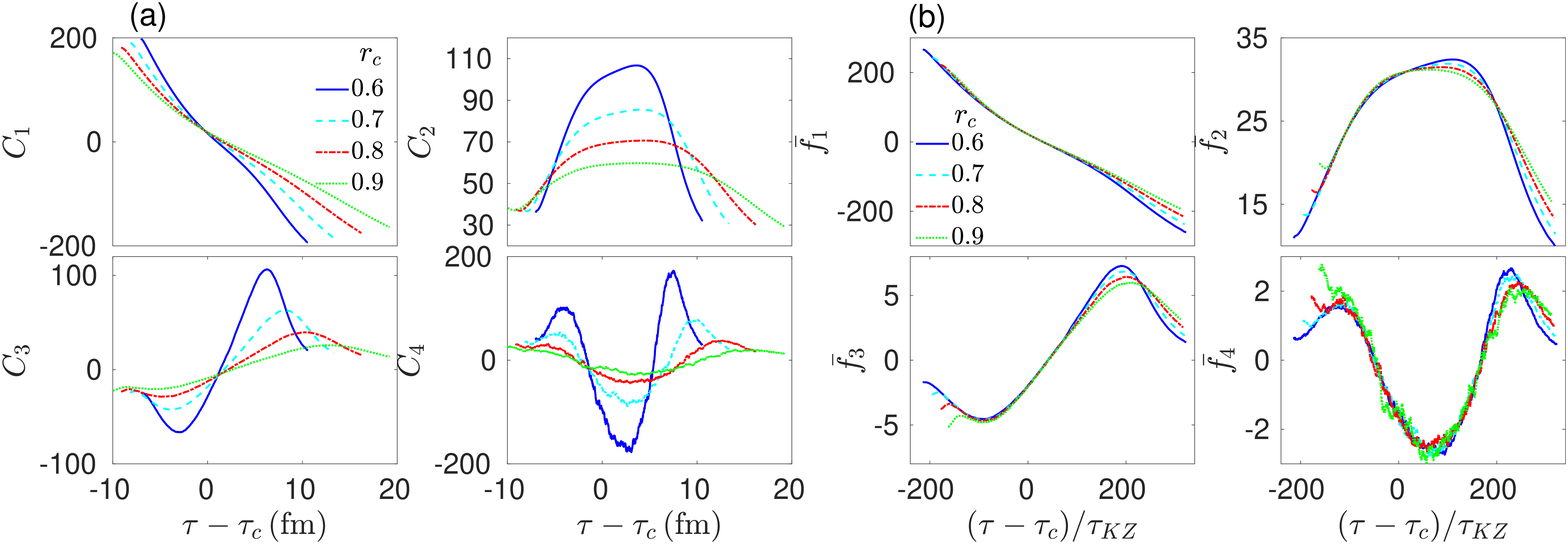}
\caption{(Color online) {Similar to Fig.~3, but evolving along type B trajectories with $r_c=0.6, \ 0.7, \ 0.8, \ 0.9$.}}
\label{B}
\end{figure*}

\begin{figure*}[tbp]
\center
\includegraphics[width=7.3in,height=2.9in]{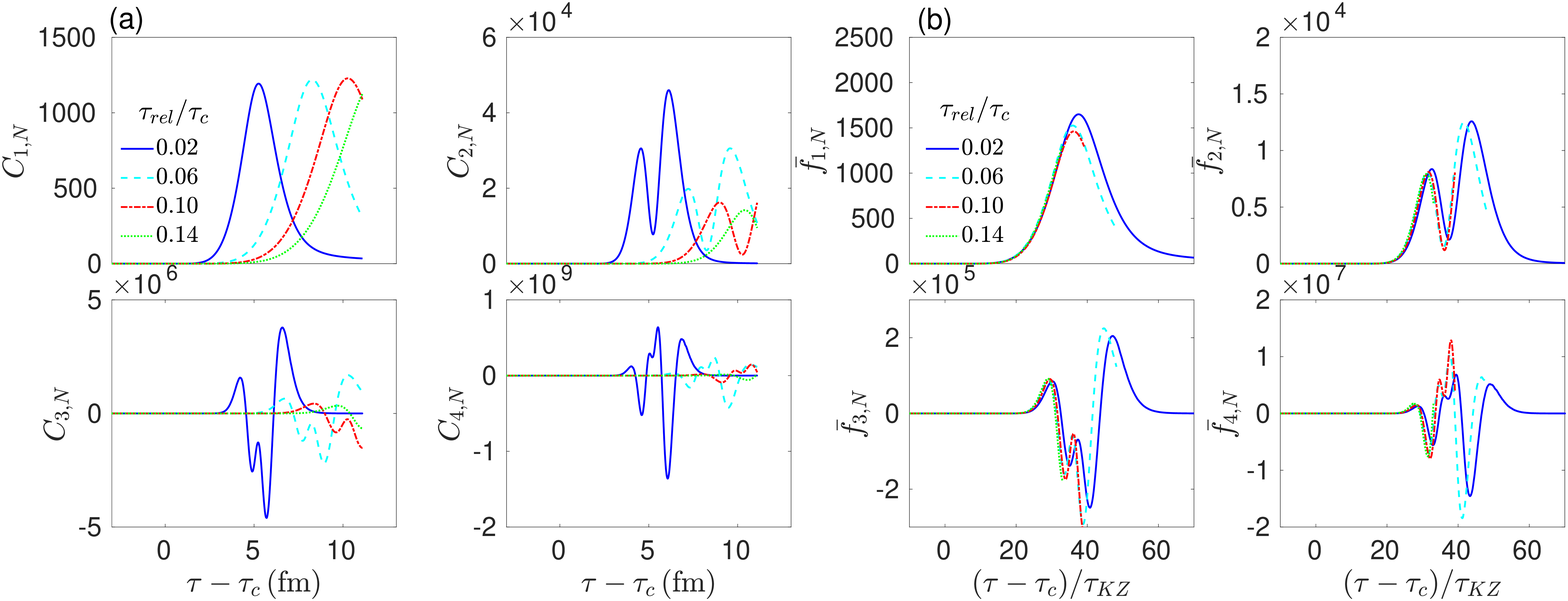}
\caption{(Color online) {(a): The cumulants $C_{n,N}\,(n=1,\dots,4)$ of net-protons as function of $\tau-\tau_c$, evolving along the trajectory of type A with $\tau_{rel}/\tau_c=0.02, \ 0.06, \ 0.10, \ 0.14$. (b): The corresponding universal functions $\bar{f}_{n,N}((\tau-\tau_{\mbox{\tiny c}})/\tau_{\mbox{\tiny KZ}},\theta_{\mbox{\tiny KZ}}) \ (n=1,\dots,4)$ as a function of  $(\tau-\tau_c)/\tau_{KZ}$.}}
\label{NetProtonA}
\end{figure*}

\begin{figure*}[tbp]
\center
\includegraphics[width=7.3in,height=2.9in]{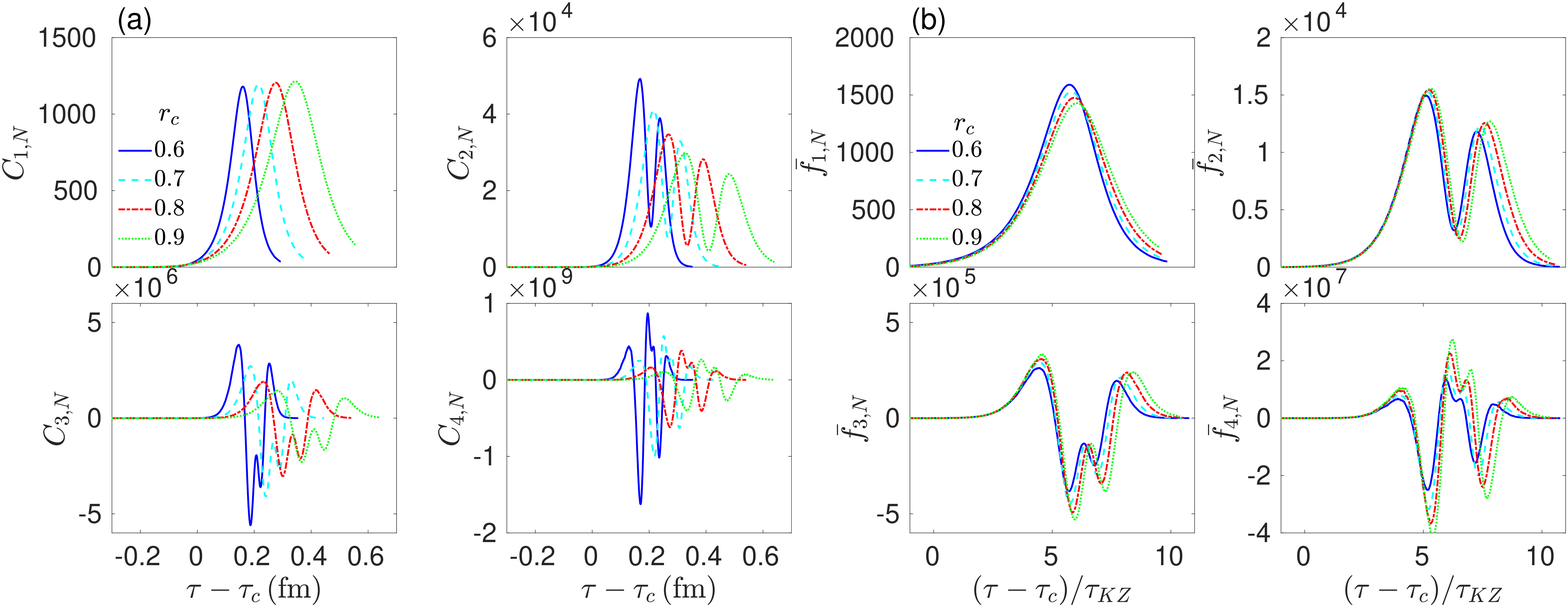}
\caption{(Color online) {Similar to Fig.~5, but evolving along type B trajectories with $r_c=0.6, \ 0.7, \ 0.8, \ 0.9$.}}
\label{NetProtonB}
\end{figure*}

In Fig.~\ref{B}, we explore the universal behavior of the sigma field with the heat bath evolving along different trajectories. For simplicity, we construct specific trajectories (type B) with approximately equal equilibrium correlation length $\xi_{eq}$ near the cross-over line, which ensures the changing rate of $\xi_{eq}$ is much smaller than the one of $\theta$. Correspondingly, the quench time can be calculated as: $\tau^\theta_{\mbox{\tiny quench}}=\left|\theta(\tau)/\partial_\tau\theta(\tau)\right|$. Similar to the above case, the proper time $\tau^*$ can be obtained from Fig.~\ref{xiKZ} (b) and the Kibble-Zurek scales $\tau_{\mbox{\tiny KZ}}$ and $l_{\mbox{\tiny KZ}}$ are calculated from Eq.~(\ref{KZ}), with which the universal functions $\bar{f}_n$ can be constructed from Eq.~(\ref{scalansatz}). In general, $\theta_{\mbox{\tiny KZ}}$ is a non-universal factor which strongly depends on
the evolving trajectories.  Here, we specifically tune the free parameter $\tau_{\mbox{\tiny rel}}/\tau_c$ to
ensure $\theta_{\mbox{\tiny KZ}}$ is a constant ($\theta_{\mbox{\tiny KZ}}=0.1$) for
these different trajectories of type B. In this case, we focus on investigating the universal scaling for such specific type of trajectories. Fig.~\ref{B} (a) show that various cumulants of the sigma field are very sensitive to the evolving trajectory. Fig.~\ref{B} (b) show that, after rescaling $C_n$ and $\tau-\tau_c$ with $l^{[-1+5(n-1)]/2}_{\mbox{\tiny KZ}}$ and $\tau_{\mbox{\tiny KZ}}$, the constructed universal functions $\bar{f}_n$ are independent on these chosen trajectories {near the critical point}.

\subsection{Approximate Kibble-Zurek scaling of net-protons}

In last subsection, we have constructed the universal functions of the sigma field that evolve with different relaxation times or with different trajectories. In this subsection, we further explore the possible universal behavior of net-protons.

In Refs.~\cite{Stephanov:2011pb,Jiang:2015hri,Ling:2015yau}, the multiplicity fluctuations of net-protons near the critical point are calculated with the modified distribution functions $f_{p/\bar{p}}(\bm{x},\bm{p})$ through coupling the protons and anti-protons with the external order parameter field. For simplicity, we take the Boltzmann distribution:
\begin{align}\label{Bolt}
f_{p/\bar{p}}(\bm{x},\bm{p}) = e^{-(E\pm\mu_B)/T}
\end{align}
where $\mu_B$ is the baryon chemical potential, the lower/upper signs are for protons/anti-protons,  and $E$ is the energy of the particle $E=\sqrt{m^2+\bm{p}^2}$. For these particles existed near the critical point,  one generally implements a variable effective mass, $m=m_0+\delta m$, to introduce critical fluctuations to the distribution function, where $m_0$ is the physical mass of the particle and $\delta m=g\sigma(\bm{x})$ comes from the interactions between the sigma field and the particles with the $\sigma NN$ coupling~\cite{Stephanov:2011pb,Jiang:2015hri,Ling:2015yau}.  In this work, we use $g=3.3,\ m_0=938.27$ MeV and set $\mu_B$ and $T$ approximatively to critical values $\mu_c=395$MeV, $T_c=160$MeV as predicted in Ref.~\cite{Fodor:2004nz}.

The total number of {net-protons} at a certain temperature $T$ and chemical
potential $\mu_B$ can be calculated through integrating $f_{p/\bar{p}}(\bm{x},\bm{p})$ over the whole phase-space:
\begin{align}\label{freezeout}
{N_{p-\bar{p}}\equiv N_p-N_{\bar{p}} = d \int \frac{d^3\bm{p}d^3\bm{x}}{(2\pi)^3}[f_p(\bm{x},\bm{p})-f_{\bar{p}}(\bm{x},\bm{p})]}
\end{align}
where the degeneracy factor $d=2$ for protons and anti-protons. With Eq.~(\ref{freezeout}), we can calculate the cumulants of net-protons and investigate the
possible Kibble-Zurek scaling near the critical point. First, we consider a simple case
with small fluctuations of the sigma field. The distribution functions $f_{p/\bar{p}}$ can be linearly
expanded as~\cite{Jiang:2015hri,Ling:2015yau}:
\begin{align}\label{linearDis}
  f_{p/\bar{p}}=f_{{p/\bar{p}},0}+ \delta f = f_{{p/\bar{p}},0}[1-g\sigma/(\gamma T)].
\end{align}
where $f_{{p/\bar{p}},0}$ is  the traditional Boltzmann distribution like the one described by Eq.~(\ref{Bolt}), but replace the variable mass $m$ by the physical mass $m_0$ of protons and anti-protons.
$\delta f$ denotes the deviation associated with the critical fluctuations
from the $\sigma NN$ coupling and $\gamma=\sqrt{m_0^2+\bm{p}^2}/m_0$.

With such expansion, various cumulants of net-protons can be calculated as:
\begin{equation}
\begin{split}
    &C_{1,N}  = \,\,\,\,\left( \int_{\bm{p}} \frac{f_{p,0}(p)-f_{\bar{p},0}(p)}{\gamma(p)}\right) \langle  \sigma \rangle,\\
    &C_{2,N} = \,\,\,\,\left( \int_{\bm{p}} \frac{f_{p,0}(p)-f_{\bar{p},0}(p)}{\gamma(p)}\right)^2 \langle (\delta \sigma)^2 \rangle, \\
    &C_{3,N} =- \left( \int_{\bm{p}} \frac{f_{p,0}(p)-f_{\bar{p},0}(p)}{\gamma(p)}\right)^3\langle(\delta \sigma)^3 \rangle,     \\
    &C_{4,N}  =\,\,\,\,\left(  \int_{\bm{p}}  \frac{f_{p,0}(p)-f_{\bar{p},0}(p)}{\gamma(p)}\right)^4[\langle (\delta \sigma)^4 \rangle - 3\langle (\delta \sigma)^2\rangle^2].
\end{split}
\end{equation}
{where the notations $\int_{\bm{p}}\equiv \frac{dg}{T}\int \frac{d^3\bm{p}}{(2\pi)^3}$ and $\delta \sigma \equiv \sigma -\langle\sigma\rangle$ .} These equations show that, with $\sigma NN$ coupling that transforms the critical fluctuations of the sigma field to the critical fluctuations of protons and anti-protons, the cumulants of the net-protons are proportional to the ones of the sigma field with the simplified liner expansion of Eq.~(\ref{linearDis}). Correspondingly, the universal scaling of net-protons behaves as the one of the sigma field as shown in Figs.~\ref{A} and~\ref{B}.

For the sigma field with large fluctuations, the linear expansion of Eq.~(\ref{linearDis}) is no longer valid. In the following calculations,  we implement the full distribution function Eq.~(\ref{Bolt})  to calculate the multiplicity fluctuations of net-protons with the configurations of the sigma field and then investigate the possible universal scaling behavior. {Fig.~\ref{NetProtonA} (a) and Fig.~\ref{NetProtonB} (a) show the time evolution of the cumulants of net-protons for a trajectory of type A with different $\tau_{rel}/\tau_c$ and for trajectories of type B with different $r_c$}. Note that {the formulae of the cumulants $C_{n,N},\,(n=1,\dots,4)$ for} net-protons  are similar to the ones of the sigma field, but replace $\sigma$ with $N_{p-\bar{p}}$ in Eqs.~(\ref{cumulants}).  As the systems evolve near the critical point, higher order cumulants of net-protons show strong oscillations and the curves associated with different evolving trajectories largely separate from each other. {In Fig.~\ref{NetProtonA} (b) and Fig.~\ref{NetProtonB}} (b), we construct the corresponding possible universal functions $\bar{f}_{n,N}((\tau-\tau_{\mbox{\tiny c}})/\tau_{\mbox{\tiny KZ}},\theta_{\mbox{\tiny KZ}}) \ (n=1,\dots,4)$ through rescaling  the cumulants of net-protons and $\tau-\tau_c$ according to Eq.~(\ref{scalansatz}). Compared with the separating / oscillating  $C_{n,N}$ curves in the left panels, the constructed $\bar{f}_{n,N}((\tau-\tau_{\mbox{\tiny c}})/\tau_{\mbox{\tiny KZ}},\theta_{\mbox{\tiny KZ}})$ approximately {converge} into one curve in Figs.~\ref{NetProtonA} (b) and \ref{NetProtonB} (b).


\section{summary and outlook}\label{summary}

In this paper, we investigated the Kibble-Zurek scaling for the critical fluctuations of the sigma field and net-protons within the framework of Langevin dynamics. We focused on two ideal cases: 1)
the systems evolve along a chosen trajectory of type A with fixed chemical potential but with different relaxation times, 2) the systems evolve along different trajectories of type B that are associated with different $r_c$ parameters.  Our event by event simulations of the Langevin dynamics demonstrated that the cumulants $C_n, ~(n=1,\dots,4)$  of the sigma field are sensitive to both the relaxation times and evolving trajectories.

Using these traditional cumulants $C_n, ~(n=1,\dots,4)$, we constructed the universal functions $\bar{f}_n((\tau-\tau_{\mbox{\tiny c}})/\tau_{\mbox{\tiny KZ}},\theta_{\mbox{\tiny KZ}})~(n=1,\dots,4)$ of the sigma field through rescaling the corresponding cumulants $C_n$ and proper time $\tau-\tau_c$ with the the characteristic scales $\tau_{\mbox{\tiny KZ}}$, $l_{\mbox{\tiny KZ}}$ and $\theta_{\mbox{\tiny KZ}}$ for the evolving systems. We found these constructed universal functions $\bar{f}_n$ are nicely  overlapped each other
in the critical regime for both case 1) and case 2), which are insensitive to the relaxation times and evolving trajectories, respectively.

For protons and anti-protons, the $\sigma NN$ coupling translate the critical fluctuations of the sigma field to the critical fluctuations of net-protons. Correspondingly, the cumulants of net-protons are sensitive to both relaxation times and evolving trajectories as the case for the sigma field. For small fluctuations of the sigma field, we found the linear expansion of the classical distribution functions $f_{p/\bar{p}}$ directly transforms the universal scaling of the sigma field to the universal scaling of net-protons. For large fluctuations of the sigma field, the numerical calculations with the full distribution functions $f_{p/\bar{p}}$ have shown that the universal behavior of net-protons are slightly broken in the critical regime, but still drastically reduce the sensitivity to the relaxation time and evolving trajectories, which even change the oscillating behavior for  higher cumulants of net-protons into an approximate universal curves.

{Finally, we emphasis that this paper focuses on investigating the universal scaling of the sigma field and net-protons for two ideal cases with specifically chosen trajectories, along which spatially uniform temperature $T$ and chemical potential $\mu$ changes  with the evolution time. These results can not be directly compared with the experimental data that involve the complex QGP fireball evolution with inhomogeneous $T(\bm{x})$ and $\mu(\bm{x})$ changing in the whole positions space. Besides, we implement the Langevin dynamics of model A to
simplify the numerical simulations, which only considers the evolution of non-conserved order parameter field near the critical point.  The multiplicity fluctuations of net-protons are introduced through $\sigma NN$ coupling in the classical distributions functions (\ref{Bolt}), which can not ensure the global charge conservation as the case in the traditional Cooper-Frye freeze-out scheme~\cite{Li:2017via,Schwarz:2017bdg}.  In the near future, such Kibble-Zurek scaling analysis should be extended to model B which directly evolves the conserved charges near the critical point. {Besides,
it is also worthwhile to develop sophisticated dynamical model near the critical point, such as hydro+, to
further investigate the possible universal scaling of the experimental observables.}}

\section*{ACKNOWLEDGEMENTS}
We would like to thank the fruitful discussion with Y.~Yin, S.~Mukherjee, M.~Stephanov,  D.~Teaney and M.~Asakawa. This work is supported by the NSFC and the MOST under Grant Nos. 11435001, 11675004 and 2015CB856900. {S.~W is also partially supported by the Beam Energy Scan Theory (BEST) Topical Collaboration during his visit to BNL}. We also gratefully acknowledge the extensive computing resources provided by the Super-computing Center of Chinese Academy of Science (SCCAS), Tianhe-1A from the National Supercomputing Center in Tianjin, China and the High-performance Computing Platform of Peking University.

\appendix

\section{Parameterization of the effective potential from 3d Ising model }

{The parameters $\sigma_0,m_\sigma,\lambda_3,\lambda_4$ in the effective potential Eq.~(\ref{effaction}) can be obtained from a mapping between the bot QCD systems and the 3d Ising model.  In the 3d Ising model, the equilibrium cumulants $M^{eq}(R,\theta),\kappa^{eq}_n(R,\theta),n=2,3,4,\dots$ can be written as~\cite{Justin:2001,Schofield:1969}:
\begin{subequations}
\begin{align}
  &M^{eq}=M_0 R^{1/3}\theta\equiv M_0a_1,\\
  &\kappa_2^{eq}=\frac{M_0}{V_4H_0} \frac{1}{R^{4/3}(3+2\theta^2)} \equiv \frac{M_0}{V_4H_0} a_2,\\
  &\kappa_3^{eq}=\frac{-M_0}{(V_4H_0)^2} \frac{4\theta(9+\theta^2)}{R^3(3-\theta^2)(3+2\theta^2)^3}\equiv\frac{-M_0}{(V_4H_0)^2} a_3,\\
  &\kappa_4^{eq}=\frac{-12M_0}{(V_4H_0)^3} \frac{81-783\theta^2+105\theta^4-5\theta^6+2\theta^8}{R^{14/3}(3-\theta^2)^3(3+2\theta^2)^5} \equiv \frac{-12M_0}{(V_4H_0)^3} a_4.
\end{align}\label{CumuEquiIsing}
\end{subequations}
{Here, $R$ and $\theta$ are the distance and angle with respect to the location of the critical point} and $V_4\equiv V/T$ (for the detail derivation of Eqs.~(\ref{CumuEquiIsing}) , please refer to Appendix A of Ref.~\cite{Mukherjee:2015swa}). {The cumulants of sigma field can also be calculated from the distribution function $P_0(\sigma)\sim \exp{(-U_0(\sigma)/T)}$, which take the forms}
\begin{align}
\begin{aligned}\label{CumuEquiQCD}
  M^{eq}=\sigma_0,\quad \kappa^{eq}_2&= \frac{\xi^2_{eq}}{V_4},\quad \kappa^{eq}_3=-\frac{2\lambda_3}{V^2_4}\xi^6_{eq},\\
  &\kappa^{eq}_{4} = \frac{6}{V^3_4}[2(\lambda_3\xi_{eq})^2-\lambda_4]\xi^8_{eq}.
\end{aligned}
\end{align}
{Comparing} Eqs.~(\ref{CumuEquiIsing}) with Eqs.~(\ref{CumuEquiQCD}) gives
\begin{align}
\begin{aligned}
  &\sigma_0(R,\theta)=M_0a_1,\quad\xi^2_{eq}(R,\theta) = 5\xi^2_{\mbox{\tiny min}}a_2,\\ &\lambda_3(R,\theta)=\frac{1}{2}\frac{H_0}{M^2_0}\frac{a_3}{a^3_2},\quad \lambda_4(R,\theta)= \frac{1}{2} \frac{H_0}{M^3_0} \frac{a^2_3}{a^5_2}+ 2\frac{H_0}{M^3_0} \frac{a_4}{a^4_2}.
\end{aligned}
\end{align}
In this work, $M_0$ and $H_0$ are two free parameters and we set $M_0=200$ MeV and $\xi_{\mbox{\tiny min}}=1$ fm.

With the linear parametric relation $r(R,\theta)= R(1-\theta^2)$,  $h(R,\theta) = R^{5/3}(3\theta-2\theta^3)$, these above $\sigma_0(R,\theta)$,~\,$\xi^2_{eq}(R,\theta)$,~\,$\lambda_3(R,\theta)$,~\,$\lambda_{4}(R,\theta)$ are converted into $\sigma_0(r,h)$,~\,$\xi^2_{eq}(r,h)$,~\,$\lambda_3(r,h)$,~\,$\lambda_{4}(r,h)$, which then can be mapped to the $T-\mu$ plane with the following linear transformation:
\begin{align}\label{map}
  \frac{T-T_c}{\Delta T} = \frac{h}{\Delta h},\quad  \frac{\mu-\mu_c}{\Delta \mu} = - \frac{r}{\Delta r},
\end{align}
where the definitions and values of the parameters  used in this paper are: $\Delta T=T_c/8,\Delta\mu=0.1\mbox{GeV},\Delta r=(5/3)^{3/4},\Delta h=1$,$T_c=0.16\,\mbox{GeV}$ and $\mu_c=0.395$ GeV.\\

\section{Analytical Kibble-Zurek scaling of model A}


In this appendix, we will analytically explain the Kibble-Zurek scaling of model A with a simplified Langevin equation, which is similar to Eq.~(\ref{Langevin3+1}), but neglect the higher order terms in the effective potential (\ref{effaction}). 
Correspondingly, the evolution equation of the sigma field after the Fourier Transformation is written as:
\begin{align}\label{LangevinFourier}
\frac{\partial \sigma(\bm{q},\tau)}{\partial \tau}= -\frac{1}{m^2_\sigma \tau_{\mbox{\tiny eff}}} \{\bm{q}^2\sigma(\bm{q},\tau)+m^2_\sigma[\sigma(\bm{q},\tau)-\sigma_0]\}+\zeta(\bm{q},\tau),
\end{align}
where the noise term in the Fourier space:

\begin{align}
&\langle \zeta(\bm{q},\tau) \rangle =0,\\
&\langle \zeta(\bm{q},\tau) \zeta(\bm{q}',\tau') \rangle =  \frac{2(2\pi)^3T}{m^2_\sigma \tau_{\mbox{\tiny eff}}} \delta^3(\bm{q}+\bm{q}')\delta(\tau-\tau').
\end{align}
With the inverse Fourier transform and the definition of Eq.(\ref{cumulants}), the evolution equation of the first and second order cumulant $C_1,\,C_2$ are written as~\footnote{The cumulants of $C_3$ and $C_4$ are zero for the simplified Langevin equation (\ref{LangevinFourier}) without higher order terms}: 
\begin{subequations}\label{EqC1C2}
\begin{align}
\frac{\partial C_1}{\partial \tau} &= -\frac{1}{ \tau_{\mbox{\tiny eff}}} [C_1-\sigma_0],\\
\frac{\partial C_2}{\partial \tau} &= -\frac{2}{ \tau_{\mbox{\tiny eff}}} \left[C_2-\frac{\xi^2_{eq}}{V_4}\right].
\end{align}
\end{subequations}
Suppose the non-universal factor is incorporated in the characteristic scales $\tau_{\mbox{\tiny KZ}},l_{\mbox{\tiny KZ}}$ and $\theta_{\mbox{\tiny KZ}}$ and the associated variables is redefined as:
\begin{align}
\tilde{\tau} \equiv(\tau-\tau_c)/\tau_{\mbox{\tiny KZ}},\,\tilde{\xi}\equiv \xi_{\mbox{\tiny eq}}/l_{\mbox{\tiny KZ}},\, \tilde{C}_1\equiv C_1/l^{-1/2}_{\mbox{\tiny KZ}},\, \tilde{C}_2\equiv C_2/l^{2}_{\mbox{\tiny KZ}}.
\end{align}
The above dynamical equation (\ref{EqC1C2}) can be rewritten as: 
\begin{subequations}\label{UnivC1C2}
\begin{align}
\frac{\partial \tilde{C}_1}{\partial \tilde{\tau}} &= -\frac{\tau_{\mbox{\tiny KZ}}}{\tau_{\mbox{\tiny rel}}  l^z_{\mbox{\tiny KZ}}(\tilde{\xi}/\xi_{\mbox{\tiny min}})^z} [\tilde{C}_1-\tilde{\sigma}_0],\\
\frac{\partial \tilde{C}_2}{\partial \tilde{\tau}} &= -2\frac{\tau_{\mbox{\tiny KZ}}}{\tau_{\mbox{\tiny rel}}  l^z_{\mbox{\tiny KZ}}(\tilde{\xi}/\xi_{\mbox{\tiny min}})^z} \left[\tilde{C}_2-\frac{\tilde{\xi}^2_{eq}}{V_4}\right],
\end{align}
\end{subequations}
Here, $V_4$ is assumed to be a constant and we use the form $\tau_{\mbox{\tiny eff}}= \tau_{\mbox{\tiny rel}}(\xi_{\mbox{\tiny eq}}/\xi_{\mbox{\tiny min}})^z$} for the effective relaxation time $\tau_{\mbox{\tiny eff}}$. The above Eqs.~(\ref{UnivC1C2}) show that, if one could eliminate the non-universal factor  $\tau_{\mbox{\tiny KZ}}/(\tau_{\mbox{\tiny rel}}  l^z_{\mbox{\tiny KZ}} )$, $\tilde{C}_1$ and $\tilde{C}_2$ as a function of $\tilde{\tau}$  become universal respectively. {In the following part of this section, we will show that the typical definition of  $\tau_{\mbox{\tiny KZ}}$ within the framework of KZM will ensure the universality of $\tilde{C}_1$ and $\tilde{C}_2$ for both Type A and Type B trajectories.}
  
For the trajectory of Type A with $r\simeq 0,\theta\simeq \pm 1$, one finds that
$\xi_{\mbox{\tiny eq}} \sim t^{-2/5}$ from Eqs.~(\ref{CumuEquiIsing}) ,(\ref{map}) and (\ref{Hubble}). Therefore, 
\begin{align}
\tau^\xi_{\mbox{\tiny quench}}=\left|\frac{\xi_{\mbox{\tiny eq}}}{\partial_\tau \xi_{\mbox{\tiny eq}}}\right|\sim\frac{5}{2} t,
\end{align}
where $t\equiv \tau-\tau_c$. As defined in Eqs.~(\ref{KZ}) for these characteristic scales 
\begin{align}\label{KZcondition}
\tau_{\mbox{\tiny KZ}} = \tau^\xi_{\mbox{\tiny quench}} = \tau_{\mbox{\tiny eff}} = \tau_{\mbox{\tiny rel}} \left(\frac{\xi_{\mbox{\tiny eq}}}{\xi_{\mbox{\tiny min}}}\right)^z,
\end{align}
Then, one could obtain
\begin{align}\label{KZnon-universal}
\tau_{\mbox{\tiny KZ}}=\tau_{\mbox{\tiny rel}}  l^z_{\mbox{\tiny KZ}},
\end{align}
which just eliminate the non-universal factor in Eq.~(\ref{UnivC1C2}) {near the critical point}.

For the trajectory of Type B, the equilibrium correlation length varies slowly, and $\tau^\xi_{\mbox{\tiny quench}}$ is large. With $\theta \sim \tau-\tau_c=t$, the quench time is:
\begin{align}
\tau^\theta_{\mbox{\tiny quench}} = \left|\frac{\theta}{\partial_\tau \theta}\right| \sim t,
\end{align}
and with the same condition (\ref{KZcondition}), we obtained the the same relation described by Eq.(\ref{KZnon-universal}), which eliminate the non-universal factor in Eq.~(\ref{UnivC1C2}).

\end{document}